\documentclass[galaxies,review,accept,oneauthor]{mdpi} 

\usepackage{graphicx,amssymb,amsmath}
\newcommand{\pd}[3]{\frac{\partial^{#3} #1}{\partial {#2}^{#3}}} 
\newcommand{\td}[3]{\frac{\mathrm{d}^{#3} #1}{\mathrm{d} {#2}^{#3}}} 
\renewcommand{\v}[1]{\ensuremath{\mathbf{#1}}} 
\newcommand{\gv}[1]{\ensuremath{\mbox{\boldmath$ #1 $}}} 
\renewcommand{\bar}[1]{\ensuremath{\overline{#1}}}
\usepackage{upgreek}
\firstpage{1} 
\pubvolume{xx}
\issuenum{1}
\articlenumber{5}
\pubyear{2019}
\copyrightyear{2019}
\history{Received: date; Accepted: date; Published: date}

\Title{Radio-Frequency Searches for Dark Matter in Dwarf~Galaxies} 

\Author{{Geoff Beck} \orcidA}

\AuthorNames{Geoff Beck}

\address[1]{%
 School of Physics, University of the Witwatersrand, Private Bag 3, {Johannesburg}, WITS-2050, South Africa; {geoffrey.beck@wits.ac.za} or geoff.m.beck@gmail.com} 

\abstract{Dwarf spheroidal galaxies have long been discussed as optimal targets for indirect dark matter searches. However, the majority of such studies have been conducted with gamma-ray instruments. In   this review, we discuss the very recent progress that has been made in radio-based indirect dark matter searches. We look at existing work on this topic and discuss the future prospects that motivate continued work in this newly developing field that promises to become, in the light of the up-coming Square Kilometre Array, a prominent component of the hunt for dark matter.}


\begin{document}


\section{Introduction}
Dwarf spheroidal galaxies (dSphs) have long been known as highly Dark Matter (DM) dominated objects with little baryonic emission that would obscure indirect detection efforts~\cite{dsph_review}. This has lead to extensive searches for DM annihilation resulting in gamma-ray emissions with numerous telescopes. Early efforts focussed on the Draco dwarf \cite{draco_2002,draco_cactus_2006} but later campaigns using the Fermi Large Area Telescope~\cite{fermi-docs} (Fermi-LAT) \cite{fermi_dsph_2012,Fermidwarves2015,fermi_ret2_2015,fermi_dsph_2016,fermi_dsph_2018}, the High Energy Stereoscopic System (HESS) \cite{hess_dsph_2007,hess_dsph_2011,hess_dsph_2014,hess_dsph_2018}, and the High Altitude Water Cherenkov (HAWC) experiment~\cite{hawc_dsph_2018} have greatly expanded the search to many other dwarf galaxy objects. However, radio continuum is another region of the spectra that few dSphs are detected in \cite{klein1992}. This suggests another possible avenue for hunting DM indirect emissions. For WIMP models, with masses above a few GeV~\cite{kolb1990}, this emission would have to be in the form of long-lived leptons emitting synchrotron radiation. Therefore, the need to disentangle the magnetic field and DM contributions to this putative emission is a complicating factor that is not present in gamma-ray detection experiments. In addition, diffusion of the emitting electrons may substantially impact expected synchrotron emissions. To characterise the diffusive environment, we need detailed information about the diffuse baryon content and turbulent magnetic field structure within the dwarf galaxy~\cite{galprop}. This is, of course, considerably complicated by the low levels of expected diffuse baryonic content and weak emissions from target objects. These complications explain the historical preference for hunting indirect DM emission in gamma-rays. However, radio instruments have several points in their favour. Firstly, their angular resolution is vastly superior to that of gamma-ray experiments, especially when interferometry is employed (\citet{fermi-docs} vs. \citet{jvla}, for instance). This is important as it can be used to avoid the confusion of diffuse dark matter emission with that of unresolved point sources. In compliment to this, radio interferometers are entering a golden age of increasing sensitivity as embodied, in the GHz frequency range, by the Jansky Very Large Array (JVLA)~\cite{jvla}, and the up-coming Square Kilometre Array (SKA)~\cite{ska2012} and its precursor experiments MeerKAT~\cite{meerkat} and the Australian Square Kilometre Array Pathfinder (ASKAP)~\cite{askap}. In addition, lower frequency experiments such as the LOw Frequency ARray (LOFAR)~\cite{lofar} and the low-frequency SKA component are pushing the boundaries of minimum detectable fluxes to levels below $1$ $\upmu$Jy. This is very promising for the indirect detection of DM as the advancement of radio astronomy techniques and technology will begin to overcome the traditional obstacles in the way of radio-based searches, allowing the strengths of radio instruments to make their impact on the hunt for DM. 

This review   therefore covers the progress that has been made towards the observation of diffuse radio emissions from dwarf spheroidal galaxies and the use of this to probe the parameter space of particle dark matter that produces electrons/positrons through annihilation or decay processes. The use of radio observation for indirect DM detection was prominently advocated in \cite{Colafrancesco2007}, using the magnetic field estimates on dSphs by \citet{klein1992} as motivation, with the Draco dwarf galaxy particularly in mind. This work spurred further searches with the Green Bank Telescope (GBT)~\cite{spekkens2013,natarajan2013} covering the Wilman I, Ursa Major II, and Coma Berenices. For this experiment, being a single dish, an external source catalogue was necessary (the NVSS was used~\cite{nvss}) to remove the contribution of point sources to the radio continuum map of the target objects. The authors of \cite{spekkens2013,natarajan2013} noted the strong dependence of their results on the magnetic field scenario within the dSphs observed, a particular issue as the instruments used could not discern the magnetic field structure. This problem of source extraction can be obviated by using interferometers to make the radio maps of the dwarf galaxy in the first place. This was the approach that was followed in the subsequent works addressing three classical and three ultra-faint dSphs with the Australian Compact Telescope Array (ATCA)~\cite{atca_I,atca_II,atca_III}. Higher achievable sensitivities could also have allowed for magnetic field estimation; however, it was found that ATCA was not sensitive enough to detect $\upmu$G level fields via rotation measure, and polarimetry is complicated by low levels of dust and gas~\cite{atca_II}. A follow-up observation was performed more recently targeting the Reticulum II dSph~\cite{regis2017}, following interest in this target by the gamma-ray DM community~\cite{fermi_ret2_2015}. The direct use of interferometers allowed for greatly improved constraints on the DM annihilation cross-section over a wide mass range. Which are, in some cases, competitive with those obtained by a Fermi-LAT study of 15 dSph objects~\cite{Fermidwarves2015}.

The results of these existing searches are   presented here and compared to a literature benchmark of the Fermi-LAT dSph searches~\cite{Fermidwarves2015}. In addition, we   follow \citet{atca_III} and present the future prospects of these radio searches by using estimation of sensitivity gains over ATCA by instruments such as {JVLA, ASKAP, MeerKAT, and the SKA. } 

This review is structured as follows: Section~\ref{sec:emm} covers all the theoretical details needed to model the synchrotron emission from electrons resulting from DM annihilation/decay, including the handling of diffusion within the radio searches presented here. In Section~\ref{sec:search}, we go into detail on the approach taken to the deep radio searches, as well as instrumental details used, in \cite{spekkens2013,natarajan2013,atca_III, regis2017}. In Section~\ref{sec:res}, we discuss the results of these aforementioned searches and compare them to our literature benchmark. Finally, in~Section~\ref{sec:prospect}, we discuss the future prospects for deep radio searches with up-coming experiments and summarise the outlook in Section~\ref{sec:out}.
 
\section{Radio Emissions from Dark Matter}
\label{sec:emm}

In this section, we cover all   necessary theoretical considerations needed to model potential radio emission that results from
DM annihilation or decay.

\subsection{Electron Source Functions from DM Annihilation/Decay}
In general, we describe the production of some particle species $i$, via DM annihilation or decay, with a source function $Q$. This $Q$ function gives the number of particles of type $i$ produced per unit volume per unit time per unit energy. This function will depend upon both the position within the DM halo $r$ and the energy of product $i$ particles, $E$. 

For annihilation, this is given by
\begin{equation}
Q_{i,A} (r,E) = \langle \sigma V\rangle \sum\limits_{f}^{} \td{N^f_i}{E}{} B_f \mathcal{N}_{\chi} (r) \; ,
\end{equation}
where $\langle \sigma V\rangle$ is the velocity-averaged DM annihilation cross-section at $0$ K, the index $f$ labels the states produced by annihilation with branching ratios $B_f$ and $i$ particle production spectra $\td{N^f_i}{E}{}$, $M_\chi$ is the WIMP mass, and finally $\mathcal{N}_{\chi} (r) = \frac{\rho_{\chi}^2}{M_\chi^2}$ is the DM particle pair density at a given halo radius $r$.

The source function in the case of DM decay is given by
\begin{equation}
Q_{i,D} (r,E) = \Gamma \sum\limits_{f}^{} \td{N^f_i}{E}{} B_f n_{\chi} (r) \; ,
\end{equation}
where $\Gamma$ is the decay rate of the DM particle, the spectra $\td{N^f_i}{E}{}$ will match those used above but for annihilation cases where the DM particle mass is half of that used for studying decay processes,  and~$n_{\chi} (r) = \frac{\rho_{\chi}}{M_\chi}$ is the DM particle number density at a given halo radius $r$.

\subsection{DM Halos of Dwarf Spheroidal Galaxies}

There are three considered density profiles that are used in deep radio searches examined in this review. They are detailed as a function of the radial coordinate $r$ below:
\begin{equation}
\begin{aligned}
\rho_N(r)=\frac{\rho_s}{\frac{r}{r_s}\left(1+\frac{r}{r_s}\right)^{2}} \; ,\\
\rho_{B} (r) = \frac{\rho_s}{\left(1 + \frac{1.52 r}{r_s}\right)\left(1+\left(\frac{1.52 r}{r_s}\right)^2\right)} \; ,\\
\rho_E(r)= \frac{1}{4}\rho_s \mbox{e}^{-\frac{2}{\alpha}\left(\left(\frac{r}{r_s} \right)^\alpha - 1\right)} \; ,
\end{aligned}
\label{eq:nfw}
\end{equation}
where $\rho_s$ is characteristic density, which normalises the density profile to virial mass of the halo $M_{vir}$; $r_s$~is the scale radius, related to the virial radius via $r_{vir} = r_s c_{vir}$ where $c_{vir}$ is the virial concentration parameter; and $\alpha$ is the free Einasto parameter. In Equation~(\ref{eq:nfw}), these density profiles are, in order, Navarro--Frenk--White (NFW)~\cite{nfw1996}, the Burkert profile~\cite{Burkert:1995}, and the Einasto profile~\cite{einasto1968}. Several profiles need to be 
considered when studying dwarf galaxies as there is some uncertainty as to their halo structure in the literature~\cite{walker2009,adams2014}.

\subsection{Diffusion of Secondary Electrons}
\label{sec:diff}
It is vital in the discussion of DM-induced radio emission to consider the diffusion and energy-loss experienced by resultant electrons. This is because both the position and energy distributions of DM-produced electrons will influence the subsequent synchrotron emission. Particularly, it has been shown in \cite{Colafrancesco2006,Colafrancesco2007,gsp2015} that the effect of diffusion on the emitted flux is highly significant in small structures such as dwarf galaxies.
The diffusion equation for electrons within the halo is given by
\begin{equation}
\begin{aligned}
\pd{}{t}{}\td{n_e}{E}{} =  \; \gv{\nabla} \left( D(E,\v{x})\gv{\nabla}\td{n_e}{E}{}\right) + \pd{}{E}{}\left( b(E,\v{x}) \td{n_e}{E}{}\right) + Q_e(E,\v{x}) \; , \label{eq:elec}
\end{aligned}
\end{equation}
where $\td{n_e}{E}{}$ is the electron spectrum, the spatial diffusion is characterised by $D(E,\v{x})$, while $b(E,\v{x})$ specifies the rate of energy-loss and $Q_e(E,\v{x})$ is the electron source function from DM annihilation or decay. The solution that is sought from such an equation is the stable equilibrium electron distribution. Two main approaches exist in the literature, one being Crank--Nicolson method for discretising derivatives. This approach is used in publicly available cosmic-ray transport codes such as DRAGON and GALPROP~\cite{dragon,galprop} and is employed in   \cite{atca_III,regis2017}. To implement this method, the time derivative is discretised
\begin{equation}
\pd{}{t}{}\td{n_e}{E}{} = \frac{\td{n_i}{E}{}(t+\Delta t) - \td{n_i}{E}{}(t)}{\Delta t} \; ,
\end{equation} 
where $i$ indicates a position $r$ (assuming spherical symmetry) and we drop the $e$ subscript for clarity. The Crank--Nicolson scheme is then
\begingroup\makeatletter\def\f@size{8}\check@mathfonts
\def\maketag@@@#1{\hbox{\m@th\normalsize\normalfont#1}}%
\begin{equation}
\frac{\td{n_i}{E}{} (t+\Delta t) - \td{n_i}{E}{}(t)}{\Delta t} = \frac{\alpha_1 \td{n_{i-1}}{E}{}(t+\Delta t) - \alpha_2 \td{n_i}{E}{}(t+\Delta t) + \alpha_3 \td{n_{i+1}}{E}{}(t+\Delta t)}{2\Delta t} - \frac{\alpha_1 \td{n_{i-1}}{E}{}(t) - \alpha_2 \td{n_i}{E}{}(t) + \alpha_3 \td{n_{i+1}}{E}{}(t)}{2\Delta t} + Q_i  \; .
\end{equation} 
\endgroup

The $\alpha$ coefficients for the $r$ discretisation are defined to match the form of Equation~(\ref{eq:elec})
\begin{align}
\frac{\alpha_1}{\Delta t} & = \left[-\frac{D + \pd{D}{r}{}}{\Delta r} + \frac{D}{\Delta r^2}\right]\Bigg\vert_{r = r_i} \; , \\
\frac{\alpha_2}{\Delta t} & = \frac{2 D(r=r_i)}{\Delta r^2} \; , \\
\frac{\alpha_3}{\Delta t} & = \left[\frac{D + \pd{D}{r}{}}{\Delta r} + \frac{D}{\Delta r^2}\right]\Bigg\vert_{r = r_i} \; .
\end{align}

The energy derivatives are discretised with the coefficients
\begin{align}
\frac{\alpha_1}{\Delta t} & = \frac{b_i(E)}{\Delta E}\; , \\
\frac{\alpha_2}{\Delta t} & = \frac{b_i(E+\Delta E)-b_i(E-\Delta E)}{\Delta E} + 2\; , \\
\frac{\alpha_3}{\Delta t} & = -\frac{b_i(E)}{\Delta E}\; .
\end{align}

The energy-loss function for inverse-{C}ompton scattering of CMB photons and synchrotron emission is~\cite{atca_II}
\begin{equation}
b(E,r) = 2.7 \times 10^{-17} \mbox{GeV s}^{-1} \left( 1 + 0.095 \left(\frac{B (r)}{\mu\mbox{G}}\right)^2\right)\left(\frac{E}{\mbox{GeV}}\right)^2 \; .
\label{eq:loss-full}
\end{equation}

The diffusion function is~\cite{atca_II}
\begin{equation}
D(E,r) = D_0 \left(\frac{B(r)}{1 \mu\mbox{G}}\right)^{-\alpha} \left(\frac{E}{1 \mbox{GeV}}\right)^{\alpha}  \; , \label{eq:diff-full}
\end{equation}
where $\alpha$ is the slope of the magnetic field power spectrum and $D_0$ ranges between $10^{28}$ and $10^{30}$~cm$^2$~s$^{-1}$.

The other approach to solving Equation~(\ref{eq:elec}) employs a semi-analytical formalism via the use of Green's functions, as used in \cite{Colafrancesco2006,Colafrancesco2007,gsp2015}.    
In this approach, it is assumed that the DM halo, and accompanying baryon distributions, have spherical symmetry. Additionally, it is assumed that the energy-loss and diffusion processes have no spatial dependence. Under these assumptions, the solution to diffusion equation takes the form
\begin{equation}
\td{n_e}{E}{} (r,E) = \frac{1}{b(E)}  \int_E^{M_\chi} d E^{\prime} \, G(r,E,E^{\prime}) Q (r,E^{\prime}) \; ,
\end{equation}
with $G(r,E,E^{\prime})$ being a Green's function. This is expressed as
\begin{equation}
\begin{aligned}
G(r,E,E^{\prime}) = & \frac{1}{\sqrt{4\pi\Delta v}} \sum_{n=-\infty}^{\infty} (-1)^n \int_0^{r_h} d r^{\prime} \; \frac{r^{\prime}}{r_n} \\ & \times \left( \exp\left(-\frac{\left(r^{\prime} - r_n\right)^2}{4\Delta v}\right) - \exp\left(-\frac{\left(r^{\prime} + r_n\right)^2}{4\Delta v}\right) \right)\frac{Q(r^{\prime})}{Q(r)} \; ,
\end{aligned}
\end{equation}
where $r_h$ is the maximum radius considered for spatial diffusion, $r_n = (-1)^n r + 2 n r_h$ are the image charge positions, and 
\begin{equation}
\Delta v =  v(u(E)) - v(u(E^{\prime})) \; ,
\end{equation}
with
\begin{equation}
\begin{aligned}
v(u(E)) = & \int_{u_{min}}^{u(E)} dx \; D(x) \; , \\
u (E) = & \int_E^{E_{max}} \frac{dx}{b(x)} \; . \\ 
\end{aligned}
\end{equation}

These last equations constitute a similar change of variables to those used in \cite{baltz1999,baltz2004} to solve Equation~(\ref{eq:elec}).
Since we have assumed that diffusion and energy-loss do not depend on halo position $r$, we include their effects via average values for the field strength and thermal plasma density. These~average values are defined as follows: $\overline{B} \equiv \sqrt{\langle B(r)^2 \rangle}$ and $\overline{n} \equiv \langle n(r) \rangle$, where the angular brackets indicate a radial average.
We can then express the spatial diffusion coefficient in terms of these averages as~\cite{Colafrancesco1998}
\begin{equation}
D(E) = D_0\left(\frac{\overline{B}}{1 \mu\mbox{G}}\right)^{-\frac{1}{3}} \left(\frac{E}{1 \mbox{GeV}}\right)^{\frac{1}{3}}  \; , \label{eq:diff}
\end{equation}
where the turbulence has been assumed Kolmogorov distributed, and $D_0$ is the diffusion constant. Note that the radial dependence of the diffusion coefficient is very weak. This justifies the assumption that we can make use of only the averaged value of the magnetic field in the diffusion coefficient. The~general electron energy-loss function is then
\begin{equation}
\begin{aligned}
b(E) = & b_{IC} E^2 (1+z)^4 + b_{sync} E^2 \overline{B}^2 \\&\; + b_{coul} \overline{n} (1+z)^3 \left(1 + \frac{1}{75}\log\left(\frac{\gamma}{\overline{n} (1+z)^3}\right)\right) \\&\; + b_{brem} \overline{n} (1+z)^3 \left( \log\left(\frac{\gamma}{\overline{n} (1+z)^3 }\right) + 0.36 \right)
\end{aligned}
\label{eq:loss}
\end{equation}
where $\gamma = \frac{E}{m_e c^2}$, $\overline{n}$ is given in cm$^{-3}$ and $b_{IC}$, $b_{synch}$, $b_{coul}$, and $b_{brem}$ are the inverse-Compton, synchrotron, Coulomb, and Bremsstrahlung energy loss factors, taken to be $0.25$, $0.0254$, $6.13$, and $1.51$, respectively, in units of $10^{-16}$ GeV s$^{-1}$. The energy $E$ is expressed in GeV and the B-field is in terms of~$\upmu$G.

\subsection{Synchrotron Emission}
An electron of energy $E$, gyrating within a  magnetic field of strength $B$ produces synchrotron emission with frequency dependent power given by~\cite{longair1994}:
\begin{equation}
P_{synch} (\nu,E,r,z) = \int_0^\pi d\theta \, \frac{\sin{\theta}^2}{2}2\pi \sqrt{3} r_e m_e c \nu_g F_{synch}\left(\frac{\kappa}{\sin{\theta}}\right) \; ,
\label{eq:power}
\end{equation}
where $\nu$ is the observed frequency, $z$ is the source redshift, the mass of the electron is given by $m_e$, the non-relativistic gyro-frequency of the electron is $\nu_g = \frac{e B}{2\pi m_e c}$, and $r_e = \frac{e^2}{m_e c^2}$ is the classical radius of the electron. Finally, $\kappa$ and $F_{synch}$ are defined as
\begin{equation}
\kappa = \frac{2\nu (1+z)}{3\nu_0 \gamma^2}\left(1 +\left(\frac{\gamma \nu_p}{\nu (1+z)}\right)^2\right)^{\frac{3}{2}} \; ,
\end{equation}
and
\begin{equation}
F_{synch}(x) = x \int_x^{\infty} dy \, K_{5/3}(y) \simeq 1.25 x^{\frac{1}{3}} \mbox{e}^{-x} \left(648 + x^2\right)^{\frac{1}{12}} \; .
\end{equation}

The synchrotron radiation emissivity at a position $r$ within the halo can then be found to be
\begin{equation}
j_{synch} (\nu,r,z) = \int_{m_e}^{M_\chi} dE \, \left(\td{n_{e^-}}{E}{} + \td{n_{e^+}}{E}{}\right) P_{synch} (\nu,E,r,z) \; .
\label{eq:emm}
\end{equation}

This quantity is the basic ingredient in determining the flux seen by a distant observer. The flux density spectrum emitted within a radius $r$ of the halo centre is found via 
\begin{equation}
S_{synch} (\nu,z) = \int_0^r d^3r^{\prime} \, \frac{j_{synch}(\nu,r^{\prime},z)}{4 \pi D_L^2} \; ,
\label{eq:flux} 
\end{equation}
where $D_L$ is the luminosity distance from observer to halo. Then, the azimuthally averaged surface brightness is given by
\begin{equation}
I_{synch} (\nu,\Theta,\Delta\Omega,z) = \int_{\Delta\Omega} d\Omega \, \int_{l.o.s} dl \, \frac{j_{synch}(\nu,l,z)}{4 \pi} \; , 
\label{eq:sb}
\end{equation}
where the integration regions $\Delta\Omega$ and $l.o.s$ define a cone of solid angle $\Delta \Omega$ around the line of sight ($l.o.s$). Note that this $l.o.s$ makes an angle $\Theta$ with the central axis of the halo. 

\section{Deep Radio Searches for Dark Matter Emissions}
\label{sec:search}

Dark matter radio emissions would be the result of synchrotron radiation emitted by electrons produced in annihilation/decay processes. This emission depends upon the DM density and so will be a truly diffuse component of the radio continuum, being on the scale of the DM halo or a few arcminutes in the case of nearby dSph targets. This means that there are two important sources of uncertainty in these searches. The first is the removal of contamination by point sources and the second is the dependence on the unknown magnetic field environment within the dwarf galaxy. The~removal of point source contributions to the radio continuum data requires a fine enough resolution to resolve such objects. This means that radio interferometers form a necessary component of the hunt for diffuse radio emission. Two approaches   have been considered in the literature: The first is the use of a single dish radio telescope (GBT) and extracting the sources via the use of a source catalogue produced by interferometer surveys (NVSS for instance). The second   is to perform the observations directly with an interferometric array (ATCA is used in \cite{atca_III,regis2017}). 

In the single telescope approach of \citet{spekkens2013}, a $40.5$ deg$^2$ area of the sky is observed at $1.4$ GHz with the GBT. The field in question contained the dSphs Wilman I, Ursa Major II, and Coma Berenices. The~resolution attained is $10^{\prime}$ and NVSS catalogue~\cite{nvss} is used to subtract the unresolved contributions of point sources to the continuum emissions. The final sensitivity attained in the source subtracted maps is around $7$ mJy per beam. Limits on DM annihilation are then obtained via the surface brightness profile of the diffuse emission being compared to expected results for DM models with varying WIMP mass, halo density profile, and annihilation cross-section. The largest uncertainties in this work are the magnetic field profile and the diffusion of the synchrotron electrons (in this case the model from~\cite{Colafrancesco2006,Colafrancesco2007} was used). The DM limits are derived under the assumption of a fiducial scenario where the diffusion constant $D_0$ is taken to be $0.1$ of that for the Milky Way, following scaling from \cite{jeltema2008}, and the magnetic field is taken to be  $B \sim 1$ $\upmu$G. The authors of these works \cite{spekkens2013,natarajan2013} also showed how sensitive their limits are to the assumptions made in regards to $B$ and $D_0$.

 \citet{atca_III} and \citet{regis2017} followed a common methodology different to that used in \cite{spekkens2013,natarajan2013}. In Refs. \cite{atca_I,atca_III}, the authors used the ATCA array to target the Carina, Fornax, Sculptor, Hercules, Segue 2, and Bootes II dSphs. Two mosaic regions of $1^{\circ}$ and $0.5^{\circ}$ were chosen containing three dSphs each. These regions were observed in a 2 GHz band around a central value of 2.1 GHz in the radio continuum. The H168 and H214 ATCA configurations were used, these having compact cores (baselines less than 100 m) and one long baseline around 4 km. A region of $20$--$30$ arcminutes was targeted around each dSph for analysis with 10--17 h (varying by dSph) on these regions in question. This allowed a nominal sensitivity to 20--40 $\upmu$Jy to be reached. Data cleaning was performed via the MFCLEAN routine from Miriad~\cite{miriad}. Sources were extracted with two approaches: SExtractor~\cite{sextractor} and SFIND within Miriad. The first of these approaches detects sources via their deviation in flux relative to the local background and the second uses a false detection rate method. These two methods were then tuned to match their source catalogues to a random position variation of around $1$ arcsecond. The resulting high resolution maps have a synthesised 8 arcsecond beam size (with 10 beams per source), a confusion limit of $3$~$\upmu$Jy, and an rms noise of 30--40 $\upmu$Jy. This is significant as it implies that source confusion will not be a factor in the analysis, as it lies below the rms sensitivity attained. A second set of maps was also produced with a Gaussian taper on the scale of $15$ arcseconds. This Gaussian tapered case results in a larger $1$~arcminute synthesised beam, which will be more suited to detecting fluxes on the scale of extended DM emission. This tapering also has the consequence of making the point sources easier to extract from the visibility plane prior to Fourier inversion, leaving an rms noise of $100$ $\upmu$Jy due to confusion limitations (which are far more significant with the taper in use). The~authors of \cite{atca_III,regis2017} always presented the DM limits from the most constraining map, choosing between either the tapered case or the high resolution maps. The authors considered two additional sources of uncertainty: bandwidth smearing and clean bias. The small size of the observed frequency band was shown to result in no significant bandwidth smearing. Clean bias, resulting from incomplete UV coverage, involves flux from sources being redistributed to the noise during data cleaning. This was mitigated by following the approach suggested in \cite{prandoni2000a} and stopping the cleaning process at a residual flux three times above the rms noise. The observations in question could not discern the magnetic field, being too insensitive to observe the rotation measure and the low dust and gas content of dSphs making polarimetry extremely~challenging.   

In Refs. \cite{atca_II,atca_III}, the Crank--Nicolson diffusion model (from Section~\ref{sec:diff}) was implemented and the authors studied three diffusion schemes. The first case is an optimistic case (OPT) where there is no spatial diffusion of the DM-produced electrons and only energy-loss at injection is considered. This OPT scenario takes the DM halo of the target dwarf galaxies to have an Einasto density profile, and magnetic field strength is calculated via an assumption of local equipartition (yielding averaged values between $4$ and $8$ $\upmu$G for the studied dSphs). The second diffusion scheme is situated between an optimistic or pessimistic scenario and is called AVG or average. This case assumes a diffusion constant given by $D_0 = 2 \times 10^{28}$ cm$^2$ s$^{-1}$  with the diffusion function $D$ experiencing an exponential increase over the scale of the stellar half-light radius $r_*$. The magnetic field in the AVG case is inferred from the rate of star formation (with the correlation normalised against data for the Large Magellanic cloud), which yields field strengths between $0.4$ and $2.0$ $\upmu$G. The DM halo profile for the dSphs is assumed to be NFW in this AVG scenario. In the third pessimistic scenario (PES), the magnetic field is inferred from star formation but only by considering data for the last Gigayear of the history of each dSph target. The halo density profile is assumed to take a cored Burkert shape. In terms of diffusion functions, PES~takes $D$ to be of the same form as in the AVG case, but, with $D_0$ being smaller by two orders of magnitude. In both the AVG and PES cases, the magnetic field decays exponentially over the scale $r_*$. 

In Ref. \cite{regis2017}, the Reticulum II dSph was targeted with the ATCA telescope in a similar configuration to \cite{atca_I,atca_II,atca_III}, complemented by large angular scale data from the KAT-7 array~\cite{kat7}. These ATCA observations involved a $23.7^{\prime}$ region containing Reticulum II and attained a $10$ $\upmu$Jy rms sensitivity, as the position of Reticulum II on the sky means that galactic foregrounds are less significant. The target was observed for 30 h in a 2 GHz band centred on $2.1$ GHz. The synthesised beam is around $7.5^{\prime\prime}\times 2.0^{\prime\prime}$ in size with well-imaged structures being above $3^{\prime}$ in extent. The KAT-7 data came from 9 h with six antennae and 44 h with five antennae in a 400 MHz band centred on 1822 MHz. These data are used as a consistency check, as the lack of long-baselines means the source subtraction is not so well defined as with ATCA. Following kinematic estimates~\cite{Bonnivard:2015}, the authors employed an Einasto density profile for the DM halo, a magnetic field model assuming $B_0 = 1$ $\upmu$G with exponential decay on the scale $r_*$ and the same diffusion function as the AVG scenario above.

\section{Search Results} 
\label{sec:res}

Preliminary work was done in this field in \cite{natarajan2013,spekkens2013} motivated by arguments from \cite{Colafrancesco2006,Colafrancesco2007}. \citet{spekkens2013} targeted the dSphs Wilman I, Coma Berenices, and Ursa Major II using data from the Greenbank telescope and the NVSS catalogue. The results of this study indicate  that, for WIMPs with masses around $100$ GeV, the DM annihilation cross-section -s constrained below $10^{-25}$ cm$^3$ s$^{-1}$. These results are extended by the second study \cite{natarajan2013}, where the authors targeted only Ursa Major II using data from Greenbank telescope and excluded (at $2\sigma$ confidence level) WIMP models with $m_{\chi} = 10$ GeV annihilating directly to electrons for cross-sections $> 10^{-26}$ cm$^3$ s$^{-1}$ and those annihilating to b quarks with $m_{\chi} = 100$ GeV and $\langle \sigma \rangle > 10^{-24}$ cm$^3$ s$^{-1}$. The results in both studies \cite{natarajan2013,spekkens2013} assume  a constant magnetic field of $1$ $\upmu$G and diffusion consistent with \cite{Colafrancesco2007}) and thus of similar magnitude to the AVG scenario in~\cite{atca_II}. 

\begin{figure}[H]
	\centering
	\resizebox{0.9\hsize}{!}{\includegraphics{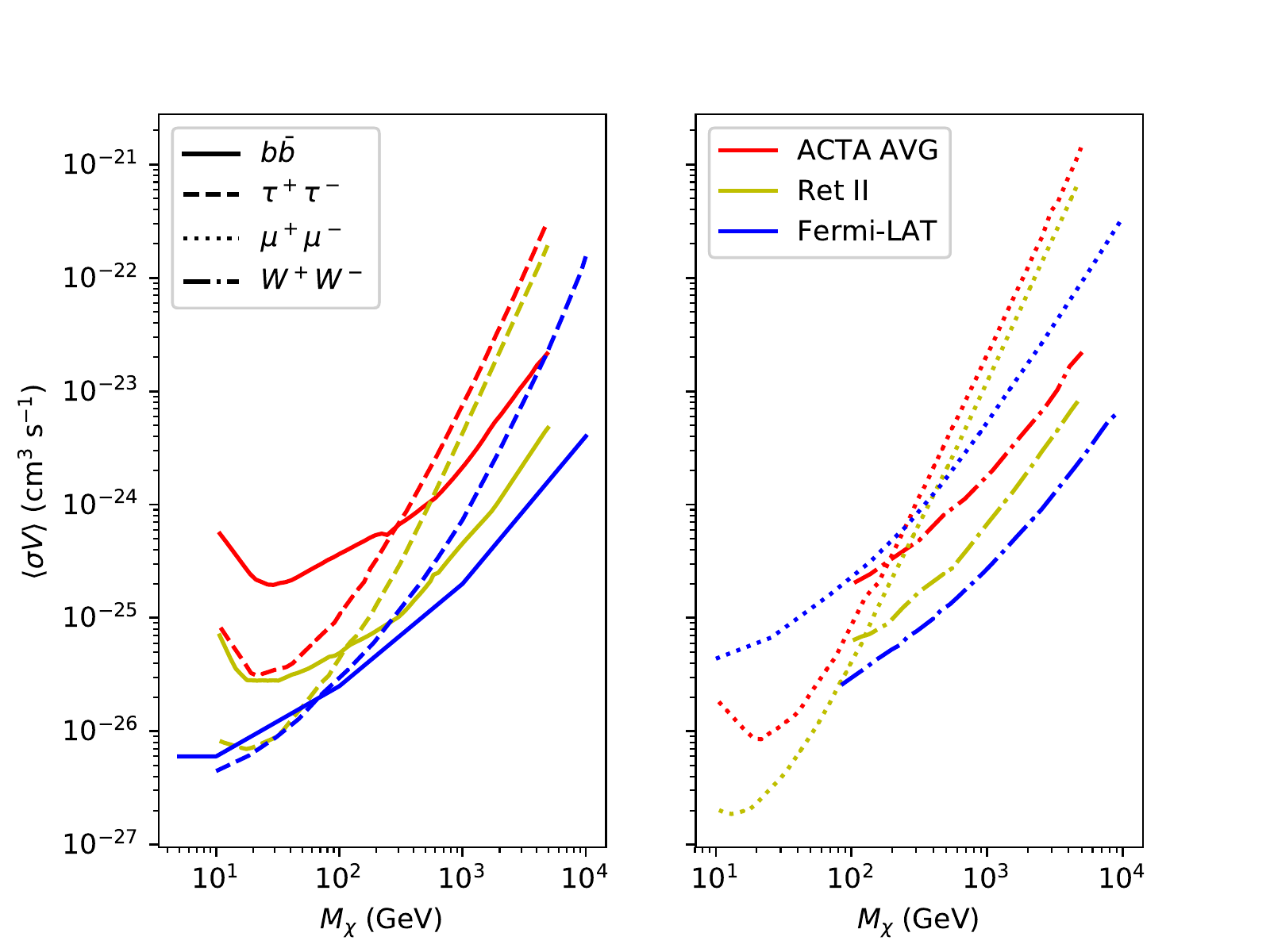}}
	\caption{Cross-section upper limits from existing searches, presented at a confidence level of $95\%$. Four annihilation channels are covered and we display results from Fermi-LAT~\cite{Fermidwarves2015}, the AVG diffusion scenario from \cite{atca_III}, and Reticulum II results from \cite{regis2017}. The left panel shows the $b$-quark and $\tau$-lepton channels. The right panel shows $W$-boson and muon channels.}
	\label{fig:compare}
\end{figure}

Now, we consider the more recent work of \cite{atca_III} (part of a trio of works \cite{atca_I,atca_II,atca_III} that contain all the observational and theoretical details of the study), where deep radio observations were performed with ATCA on the Carina, Fornax, Sculptor, Hercules Segue 2, and Bootes II dSphs. In this case ,the authors studied three diffusion schemes detailed above in Section~\ref{sec:search}. Limits on the annihilation cross-section span around six or seven orders of magnitude between the three models with the largest gap being between OPT and AVG (AVG and PES differ by around 2 orders of magnitude). For individual galaxies, in the AVG scenario, the constraints are competitive with those found in \cite{spekkens2013} over a wide mass range (10--5000 GeV). However, a combined constraint produced in the AVG is considerably stronger (as can be seen for several annihilation channels in Figure~\ref{fig:compare}). 

 \citet{regis2017} found no evidence of diffuse radio emission in Reticulum II and thus derived constraints on the WIMP annihilation cross-section or decay rate from this (following a model similar to AVG above). These constraints are displayed in Figure~\ref{fig:compare}. 

What is evident in Figure~\ref{fig:compare} is that the limits from non-observation of diffuse emission in Reticulum II from \cite{regis2017} are up to an order of magnitude stronger than those from \cite{atca_III} for all displayed annihilation channels (note that the diffusion scenarios are very similar in this comparison). Particularly, the~$b\bar{b}$ channel is an order of magnitude better in \cite{regis2017} but other channels are far more similar. When these constraints are placed into literature context against a benchmark like the Fermi-LAT dwarf galaxy gamma-ray limits~\cite{Fermidwarves2015} we find that \cite{atca_III} is around an order of magnitude less stringent than Fermi. In~the case of \cite{regis2017}, we see that these limits are more competitive with those from gamma-rays than~\cite{atca_III}, being~more stringent for low masses with the muon annihilation channel and within a order of magnitude of Fermi-LAT otherwise. It is worth noting that we are comparing a single dwarf galaxy with \cite{regis2017}, and six galaxies in \cite{atca_III}, against a combined 15 galaxy analysis in \cite{fermi-decay}, which indicates the competitive potential of the radio approach.

\begin{figure}[H]
	\centering
	\resizebox{0.9\hsize}{!}{\includegraphics{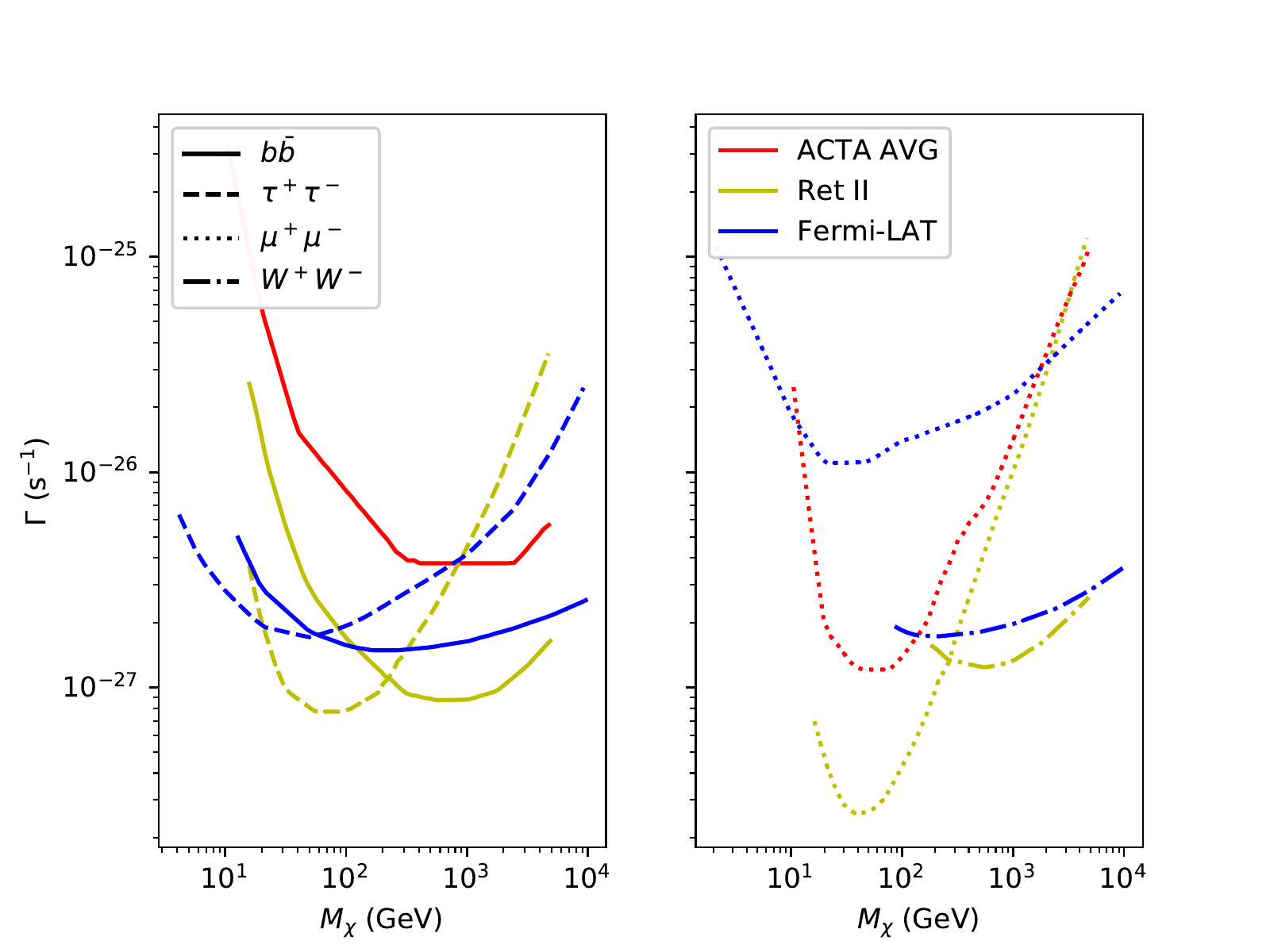}}
	\caption{Decay rate limits from existing searches, presented at a confidence level of $95\%$. Four annihilation channels are covered and we display results from Fermi-LAT~\cite{fermi-decay}, the AVG diffusion scenario from \cite{atca_III} (limited to channels quoted in the reference), and Reticulum II results from \cite{regis2017}. The left panel shows the $b$-{quark} ($b\bar{b}$) and $\tau$-lepton channels. The right panel shows $W$-boson and muon~channels.}
	\label{fig:compare_decay}
\end{figure}

In Figure~\ref{fig:compare_decay}, we display analogous results for the scenario of decaying dark matter particles. These~are compared against the Fermi-LAT dwarf galaxy limits from \cite{fermi-decay}. In this case, we plot the limited channels presented in \cite{atca_III} for this particular study. What is evident is that the limits from~\cite{regis2017} make some improvements over those from gamma-rays. This increase is more than an order of magnitude for masses below $1$ TeV in the case of the muon channel (where \cite{atca_III} is also superior to Fermi-LAT by around an order of magnitude), and factor $2$ improvements at mass between $100$~GeV and $10$ TeV for $b\bar{b}$, and similar increases between $20$ and $1000$ GeV for the tau-lepton channel. The~$W$-boson channel is very similar for both gamma-ray and radio studies. 

\section{Future Prospects}
\label{sec:prospect}

Many new generation radio observatories are either coming online presently or are expected within the near future. We   cover those that operate in a similar bandwidth to \cite{spekkens2013,natarajan2013,atca_III,regis2017} in detail but do not discuss those experiments that operate outside this frequency band as their projections are not easily comparable to the results presented by the aforementioned studies. We   note, however, that it has been argued that LOFAR~\cite{lofar} may have some potential in indirect DM detection~\cite{leite2016,storm2017}, however neither of these studies directly addressed the dSph scenario considered here.

\begin{figure}[H]
	\centering
	\resizebox{0.9\hsize}{!}{\includegraphics{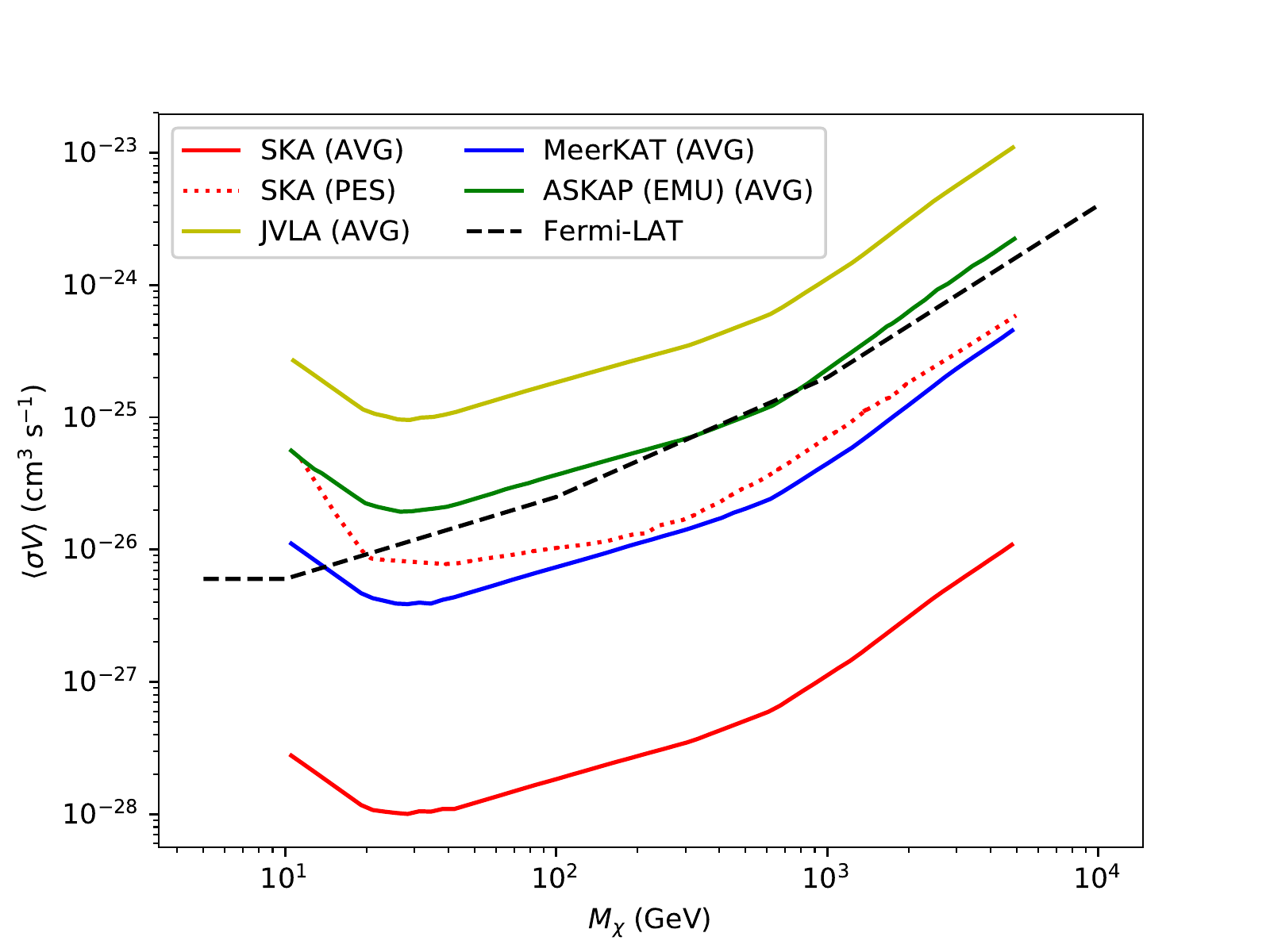}}
	\caption{Cross-section upper limit prospects for the JVLA, ASKAP EMU, MeerKAT, and SKA from~\cite{atca_III} compared to Fermi-LAT limits~\cite{Fermidwarves2015}. These limits are displayed at $95\%$ confidence interval and only for the $b$-quark ($b\bar{b}$) channel.}
	\label{fig:prospect}
\end{figure}

A particular example of improvements over the results from \cite{atca_III,regis2017} could be drawn from deep radio searches with the existing JVLA~\cite{jvla} telescope. In particular, making use of the D configuration with baselines between 1 km and 35 m to observe both the large scale diffuse emission and perform source extraction. This instrument is capable of an rms sensitivity in the GHz range of around $10$~$\upmu$Jy for 1 h per pointing which can provide a substantial advantage over the ATCA observations used in~\cite{atca_III,regis2017}. Despite this choice of optimal instrumental configuration, JVLA data would still require even longer baseline observations to remove point sources. This is because the D configuration confusion limit approaches $90$ $\upmu$Jy and will thus impact on the potential to probe faint diffuse radio fluxes. Thus, overall, the JVLA may produce as much as factor of 2 improvement on the results of \cite{atca_III,regis2017}, as shown for the \cite{atca_III} targets in Figure~\ref{fig:prospect}. Such an improvement would make the limits from Reticulum II in \cite{regis2017} very competitive with the gamma-ray case presented for a study of 15 dSphs by Fermi-LAT and shown in the same plot. 
For the SKA~\cite{ska2012} precursor ASKAP~\cite{askap}, its survey project EMU~\cite{emu} will attain GHz continuum rms sensitivity of $10$ $\upmu$Jy over a 30 square degree area with an angular resolution of $10$ arcseconds. There are $14$ known Milky-Way satellites in this area and many more dwarf galaxy detection could be expected from southern-sky surveys \cite{des,skymap,lsst} (as the Sloan Digital Sky Survey more than doubled the number of known northern-sky dSphs). The authors of \cite{atca_III} argued for an increase in sensitivity over their results of a factor of 5--10, factoring in an increase in the dSph sample observed (as can be seen in Figure~\ref{fig:prospect}). 

In the case of the SKA precursor MeerKAT~\cite{meerkat}, the field of view will be smaller than ASKAP but it has a faster survey speed and will be well suited to deep dSph observations. This is because it should potentially obtain $\lesssim$1 $\upmu$Jy rms sensitivity around the GHz range (with an integration time approaching 1000 h). As can be seen in Figure~\ref{fig:prospect}, MeerKAT can probe more of the parameter space than Fermi-LAT for the AVG diffusion scenario.

The SKA itself will achieve up to a two orders of magnitude improvement on its precursors~\cite{ska2012} sensitivity for the mid-frequency band ranging   350--1050 MHz. It will also have the advantage of being able to determine magnetic field structures via rotation measures for fields around $1$ $\upmu$G in the dSph environment. In Figure~\ref{fig:prospect}, this leads to being able to produce superior limits to Fermi-LAT (for WIMP mass $\geq 20$ GeV) even in the PES diffusion scenario. These sensitivity improvements come with caveats. The first is that, with the SKA reaching such potentially faint diffuse fluxes, confusion limits will unpredictably affect the source subtraction adjustment to the sensitivity. Secondly, the low level star formation emissions expected within the dSph could be a complicating factor for very faint fluxes. This second caveat can be mitigated through a combination of the use of optical correlations and the fine angular resolution of the SKA itself to identify star formation contributions. However, these caveats do mean that the projection in Figure~\ref{fig:prospect} is likely optimistic.

\section{Outlook}
\label{sec:out}

The preceding discussion indicates that radio searches for DM annihilation/decay in dwarf galaxies is entering into an age in which it becomes more sensitive to putative DM emissions than gamma-ray telescopes. This means that these kinds of deep radio GHz frequency searches will become a leading candidate for indirect DM hunting as they mature with the arrival of future telescopes such as the SKA and its precursors. Supplementing these GHz projects with lower frequencies via LOFAR or SKA-LOW will make the study of DM via diffuse emission in dwarf galaxies a powerful probe into one of the largest hiatuses in current models of cosmology.

\vspace{6pt} 




\funding{This research received no external funding.}

\acknowledgments{This review is dedicated to the memory of my colleague and mentor Sergio Colafrancesco.}

\conflictsofinterest{The author declares no conflict of interest.} 

\abbreviations{The following abbreviations are used in this manuscript:\\

\noindent 
\begin{tabular}{@{}ll}
DM & Dark Matter\\
dSph & Dwarf spheroidal galaxy\\
SKA & Square Kilometre Array\\
ATCA & Australian Telescope Compact Array\\
GBT & Green Bank Telescope\\
JVLA & Jansky Very Large Array\\
ASKAP & Australian Square Kilometre Array Pathfinder\\
EMU & Evolutionary Map of the Universe\\
KAT & Karoo Array Telescope\\
LOFAR & LOw Frequency ARray
\end{tabular}}

%


\reftitle{References}



\newpage
\begin{thebibliography}{999}
\providecommand{\natexlab}[1]{#1}

\bibitem[Mateo(1998)]{dsph_review}
Mateo, M.
\newblock {Dwarf galaxies of the Local Group}.
\newblock {\em Ann. Rev. Astron. Astrophys.} {\bf 1998}, {\em 36},~435--506,
\newblock
  doi:{\changeurlcolor{black}\href{https://doi.org/10.1146/annurev.astro.36.1.435}{\detokenize{10.1146/annurev.astro.36.1.435}}}.

\bibitem[Tyler(2002)]{draco_2002}
Tyler, C.
\newblock Particle dark matter constraints from the Draco dwarf galaxy.
\newblock {\em Phys. Rev. D} {\bf 2002}, {\em 66},~023509,
\newblock
  doi:{\changeurlcolor{black}\href{https://doi.org/10.1103/PhysRevD.66.023509}{\detokenize{10.1103/PhysRevD.66.023509}}}.

\bibitem[Profumo and Kamionkowski(2006)]{draco_cactus_2006}
Profumo, S.; Kamionkowski, M.
\newblock {Dark matter and the cactus gamma-ray excess from draco}.
\newblock {\em JCAP} {\bf 2006}, 
\newblock
  doi:{\changeurlcolor{black}\href{https://doi.org/10.1088/1475-7516/2006/03/003}{\detokenize{10.1088/1475-7516/2006/03/003}}}.

\bibitem[Atwood and others for~the Fermi/LAT~collaboration(2009)]{fermi-docs}
Atwood, W.B.; Abdo A. A.; Ackermann, M.; Althouse, W.; Anderson, B.; Axelsson, M.; Baldini, L.; Ballet, J.; Band, D. L.; Barbiellini, G.; {et al. for~the Fermi/LAT~collaboration.} 
\newblock{The Large Area Telescope on the Fermi Gamma-ray Space Telescope Mission}
\newblock {\em Astrophys. J.} {\bf 2009}, {\em 697},~1071--1102.
\newblock
doi:{\changeurlcolor{black}\href{https://doi.org/10.1088/0004-637X/697/2/1071}{\detokenize{10.1088/0004-637X/697/2/1071}}}.

\bibitem[Ackermann \em{et~al.}(2012)Ackermann et~al.]{fermi_dsph_2012}
Ackermann, M.; Albert, A.; Baldini, L.; Ballet, J.; Barbiellini, G.; Bastieri, D.; Bechtol, K.; Bellazzini, R.; Blandford, R.D.; Bloom, E.D.; et al.
\newblock {Search for Dark Matter Satellites using the FERMI-LAT}.
\newblock {\em Astrophys. J.} {\bf 2012}, {\em 747},~121,
\newblock
  doi:{\changeurlcolor{black}\href{https://doi.org/10.1088/0004-637X/747/2/121}{\detokenize{10.1088/0004-637X/747/2/121}}}.

\bibitem[Ackermann \em{et~al.}(2015)Ackermann et~al.]{Fermidwarves2015}
Ackermann, M.; Albert, A.; Anderson, B.; Atwood, W.B.; Baldini, L.; Barbiellini, G.; Bastieri, D.; Bechtol,~K.; Bellazzini, R.; Bissaldi, E.; et al.
\newblock Searching for Dark Matter Annihilation from Milky Way Dwarf
  Spheroidal Galaxies with Six Years of Fermi Large Area Telescope Data.
\newblock {\em Phys. Rev. Lett.} {\bf 2015}, {\em 115},~231301,
\newblock
  doi:{\changeurlcolor{black}\href{https://doi.org/10.1103/PhysRevLett.115.231301}{\detokenize{10.1103/PhysRevLett.115.231301}}}.

\bibitem[Geringer-Sameth \em{et~al.}(2015)Geringer-Sameth, Walker, Koushiappas,
  Koposov, Belokurov, Torrealba, and Evans]{fermi_ret2_2015}
Geringer-Sameth, A.; Walker, M.G.; Koushiappas, S.M.; Koposov, S.E.; Belokurov,
  V.; Torrealba, G.; Evans,~N.W.
\newblock Indication of Gamma-Ray Emission from the Newly Discovered Dwarf
  Galaxy Reticulum II.
\newblock {\em Phys. Rev. Lett.} {\bf 2015}, {\em 115},~081101,
\newblock
  doi:{\changeurlcolor{black}\href{https://doi.org/10.1103/PhysRevLett.115.081101}{\detokenize{10.1103/PhysRevLett.115.081101}}}.

\bibitem[Li \em{et~al.}(2016)Li, Liang, Duan, Shen, Huang, Li, Fan, Liao, Feng,
  and Chang]{fermi_dsph_2016}
Li, S.; Liang, Y.F.; Duan, K.K.; Shen, Z.Q.; Huang, X.; Li, X.; Fan, Y.Z.;
  Liao, N.H.; Feng, L.; Chang, J.
\newblock {Search for gamma-ray emission from eight dwarf spheroidal galaxy
  candidates discovered in Year Two of Dark Energy Survey with Fermi-LAT data}.
\newblock {\em Phys. Rev.} {\bf 2016}, {\em D93},~043518,
\newblock
  doi:{\changeurlcolor{black}\href{https://doi.org/10.1103/PhysRevD.93.043518}{\detokenize{10.1103/PhysRevD.93.043518}}}.

\bibitem[Li \em{et~al.}(2018)Li, Duan, Liang, Xia, Shen, Li, Liao, Feng, Yuan,
  Fan, and Chang]{fermi_dsph_2018}
Li, S.; Duan, K.K.; Liang, Y.F.; Xia, Z.Q.; Shen, Z.Q.; Li, X.; Liao, N.H.;
  Feng, L.; Yuan, Q.; Fan, Y.Z.; et al.
\newblock Search for gamma-ray emission from the nearby dwarf spheroidal
  galaxies with 9 years of Fermi-LAT data.
\newblock {\em Phys. Rev. D} {\bf 2018}, {\em 97},~122001,
\newblock
  doi:{\changeurlcolor{black}\href{https://doi.org/10.1103/PhysRevD.97.122001}{\detokenize{10.1103/PhysRevD.97.122001}}}.

\bibitem[Aharonian(2008)]{hess_dsph_2007}
Aharonian, F.
\newblock {Observations of the Sagittarius Dwarf galaxy by the H.E.S.S.
  experiment and search for a Dark Matter signal}.
\newblock {\em Astropart. Phys.} {\bf 2008}, {\em 29},~55--62; Erratum in {\bf 2010}, {\em 33}, ~274--275,
\newblock 
  doi:{\changeurlcolor{black}\href{https://doi.org/10.1016/j.astropartphys.2007.11.007}{\detokenize{10.1016/j.astropartphys.2007.11.007}}}. 


\bibitem[Abramowski \em{et~al.}(2011)Abramowski et~al.]{hess_dsph_2011}
Abramowski, A.; Acero, F.; Aharonian, F.; Akhperjanian, A.G.; Anton, G.; Barnacka, A.; De~Almeida,~U.B.; Bazer-Bachi, A.R.; Becherini, Y.; Becker, J.; et al.
\newblock H.E.S.S. constraints on dark matter annihilations towards the
  sculptor and carina dwarf galaxies.
\newblock {\em Astropart. Phys.} {\bf 2011}, {\em 34},~608--616,
\newblock
  doi:{\changeurlcolor{black}\href{https://doi.org/https://doi.org/10.1016/j.astropartphys.2010.12.006}{\detokenize{10.1016/j.astropartphys.2010.12.006}}}.

\bibitem[Abramowski \em{et~al.}(2014)Abramowski et~al.]{hess_dsph_2014}
Abramowski, A.; Aharonian, F.; Benkhali, F.A.; Akhperjanian, A.G.; Angüner, E.; Backes, M.; Balenderan,~S.; Balzer, A.; Barnacka, A.; Becherini, Y.; et al.
\newblock {Search for dark matter annihilation signatures in H.E.S.S.
  observations of Dwarf Spheroidal Galaxies}.
\newblock {\em Phys. Rev. D} {\bf 2014}, {\em 90},~112012,
\newblock
  doi:{\changeurlcolor{black}\href{https://doi.org/10.1103/PhysRevD.90.112012}{\detokenize{10.1103/PhysRevD.90.112012}}}.

\bibitem[Abdalla \em{et~al.}(2018)Abdalla et~al.]{hess_dsph_2018}
Abdalla, H.; Aharonian, F.; Benkhali, F.A.; Angüner, E.O.; Arakawa, M.; Arcaro, C.; Armand, C.; Arrieta, M.; Backes, M.; Barnard, M.; et al.
\newblock {Searches for gamma-ray lines and 'pure WIMP' spectra from Dark
  Matter annihilations in dwarf galaxies with H.E.S.S}.
\newblock {\em J. Cosmol. Astropart. Phys.} {\bf 2018}, doi:10.1088/1475-7516/2018/11/037.

\bibitem[Albert \em{et~al.}(2018)Albert et~al.]{hawc_dsph_2018}
Albert, A.; Alfaro, R.; Alvarez, C.; Álvarez, J.D.; Arceo, R.; Arteaga-Velázquez, J.C.; Rojas, D.A.; Solares, H.A.; Bautista-Elivar, N.; Becerril, A.; et al.
\newblock {Dark Matter Limits from Dwarf Spheroidal Galaxies with The HAWC
  Gamma-Ray Observatory}.
\newblock {\em Astrophys. J.} {\bf 2018}, {\em 853},~154,
\newblock
  doi:{\changeurlcolor{black}\href{https://doi.org/10.3847/1538-4357/aaa6d8}{\detokenize{10.3847/1538-4357/aaa6d8}}}.

\bibitem[Klein \em{et~al.}(1992)Klein, Giovanardi, Altschuler, and
  Wunderlich]{klein1992}
Klein, U.; Giovanardi, C.; Altschuler, D.R.; Wunderlich, E.
\newblock A sensitive radio continuum survey of low surface brightness dwarf
  galaxies.
\newblock {\em Astron. Astrophys.} {\bf 1992}, {\em 255},~49--58.

\bibitem[Kolb and Turner(1990)]{kolb1990}
Kolb, E.W.; Turner, M.S.
\newblock {The Early Universe}.
\newblock {\em Front. Phys.} {\bf 1990}, {\em 69},~115-152. 

\bibitem[Strong and Moskalenko(1998)]{galprop}
Strong, A.W.; Moskalenko, I.V.
\newblock {Propagation of cosmic-ray nucleons in the galaxy}.
\newblock {\em Astrophys. J.} {\bf 1998}, {\em 509},~212--228,
\newblock
  doi:{\changeurlcolor{black}\href{https://doi.org/10.1086/306470}{\detokenize{10.1086/306470}}}.

\bibitem[Perley \em{et~al.}(2011)Perley, Chandler, Butler, and Wrobel]{jvla}
Perley, R.A.; Chandler, C.J.; Butler, B.J.; Wrobel, J.M.
\newblock The Expanded Very Large Array: A New Telescope for New Science.
\newblock {\em Astrophys. J. Lett.} {\bf 2011}, {\em 739},~L1.

\bibitem[Dewdney \em{et~al.}(2012)Dewdney, Turner, Millenaar, McCool, Lazio,
  and Cornwell]{ska2012}
Dewdney, P.; Turner, W.; Millenaar, R.; McCool, R.; Lazio, J.; Cornwell, T.
\newblock SKA Baseline Design Document { 2012}. Available online:
\newblock {
  \url{http://www.skatelescope.org/wp-content/uploads/2012/07/SKA-TEL-SKO-DD-001-1_BaselineDesign1.pdf}} ({accessed on 11/11/2018}). 

\bibitem[Booth \em{et~al.}(2009)Booth, de~Blok, Jonas, and Fanaroff]{meerkat}
Booth, R.; de~Blok, W.; Jonas, J.; Fanaroff, B.
\newblock MeerKAT Key Project Science, Specifications, and Proposals.
\newblock {\em arXiv} {\bf 2009}, { arXiv:0910.2935}.

\bibitem[McConnell \em{et~al.}(2016)McConnell, Allison, Bannister, Bell,
  Bignall, Chippendale, Edwards, Harvey-Smith, Hegarty, Heywood, and
  et~al.]{askap}
McConnell, D.; Allison, J.R.; Bannister, K.; Bell, M.E.; Bignall, H.E.;
  Chippendale, A.P.; Edwards, P.G.; Harvey-Smith, L.; Hegarty, S.; Heywood, I.;
  et~al.
\newblock The Australian Square Kilometre Array Pathfinder: Performance of the
  Boolardy Engineering Test Array.
\newblock {\em Publ. Astron. Soc. Aust.} {\bf
  2016}, {\em 33},~e042,
\newblock
  doi:{\changeurlcolor{black}\href{https://doi.org/10.1017/pasa.2016.37}{\detokenize{10.1017/pasa.2016.37}}}.

\bibitem[{van Haarlem, M. P.} \em{et~al.}(2013){van Haarlem, M. P.}
  et~al.]{lofar}
{van Haarlem, M. P.}; Wise, M.W.; Gunst, A.W.; Heald, G.; McKean, J.P.; Hessels, J.W.; De Bruyn, A.G.; Nijboer,~R.; Swinbank, J.; Fallows, R.; et al.
\newblock LOFAR: The Low-Frequency Array.
\newblock {\em Astron. Astrophys.} {\bf 2013}, {\em 556},~A2,
\newblock
  doi:{\changeurlcolor{black}\href{https://doi.org/10.1051/0004-6361/201220873}{\detokenize{10.1051/0004-6361/201220873}}}.

\bibitem[Colafrancesco \em{et~al.}(2007)Colafrancesco, Profumo, and
  Ullio]{Colafrancesco2007}
Colafrancesco, S.; Profumo, S.; Ullio, P.
\newblock Detecting dark matter WIMPs in the Draco dwarf: A multi-wavelength
  perspective.
\newblock {\em Phys. Rev. D} {\bf 2007}, {\em 75},~023513.

\bibitem[Spekkens \em{et~al.}(2013)Spekkens, Mason, Aguirre, and
  Nhan]{spekkens2013}
Spekkens, K.; Mason, B.S.; Aguirre, J.E.; Nhan, B.
\newblock A Deep Search for Extended Radio Continuum Emission From Dwarf
  Spheroidal Galaxies: Implications for Particle Dark Matter.
\newblock {\em Astrophys. J.} {\bf 2013}, {\em 773},~61.

\bibitem[Natarajan \em{et~al.}(2013)Natarajan et~al.]{natarajan2013}
Natarajan, A.; others.
\newblock Bounds on Dark Matter Properties from Radio Observations of Ursa
  Major II using the Green Bank Telescope.
\newblock {\em Phys. Rev. D} {\bf 2013}, {\em 88},~083535.

\bibitem[Condon \em{et~al.}(1998)Condon, Cotton, Greisen, Yin, Perley, Taylor,
  and Broderick]{nvss}
Condon, J.J.; Cotton, W.D.; Greisen, E.W.; Yin, Q.F.; Perley, R.A.; Taylor,
  G.B.; Broderick, J.J.
\newblock {The NRAO VLA Sky survey}.
\newblock {\em Astron. J.} {\bf 1998}, {\em 115},~1693--1716,
\newblock
  doi:{\changeurlcolor{black}\href{https://doi.org/10.1086/300337}{\detokenize{10.1086/300337}}}.

\bibitem[Regis \em{et~al.}(2015{\natexlab{a}})Regis, Richter, Colafrancesco,
  Massardi, de~Blok, Profumo, and Orford]{atca_I}
Regis, M.; Richter, L.; Colafrancesco, S.; Massardi, M.; de~Blok, W.J.G.;
  Profumo, S.; Orford, N.
\newblock Local Group dSph radio survey with ATCA – I: Observations and
  background sources.
\newblock {\em Mon. Not. R. Astron. Soc.} {\bf 2015},
  {\em 448},~3731--3746,
\newblock
  doi:{\changeurlcolor{black}\href{https://doi.org/10.1093/mnras/stu2747}{\detokenize{10.1093/mnras/stu2747}}}.

\bibitem[Regis \em{et~al.}(2015{\natexlab{b}})Regis, Richter, Colafrancesco,
  Profumo, de Blok, and Massardi]{atca_II}
Regis, M.; Richter, L.; Colafrancesco, S.; Profumo, S.; de Blok, W.J.G.;
  Massardi, M.
\newblock Local Group dSph radio survey with ATCA---II. Non-thermal diffuse
  emission.
\newblock {\em Mon. Not. R. Astron. Soc.} {\bf 2015},
  {\em 448},~3747--3765,
\newblock
  doi:{\changeurlcolor{black}\href{https://doi.org/10.1093/mnras/stv127}{\detokenize{10.1093/mnras/stv127}}}.

\bibitem[Regis \em{et~al.}(2014)Regis, Colafrancesco, Profumo, de~Blok,
  Massardi, and Richter]{atca_III}
Regis, M.; Colafrancesco, S.; Profumo, S.; de~Blok, W.; Massardi, M.; Richter,
  L.
\newblock Local Group dSph radio survey with ATCA (III): Constraints on
  particle dark matter.
\newblock {\em J. Cosmol. Astropart.  Phys.} {\bf 2014}, {\em
  2014},~016.

\bibitem[Regis \em{et~al.}(2017)Regis, Richter, and Colafrancesco]{regis2017}
Regis, M.; Richter, L.; Colafrancesco, S.
\newblock Dark matter in the Reticulum II dSph: A radio search.
\newblock {\em J. Cosmol. Astropart.  Phys.} {\bf 2017}, {\em
  2017},~025.

\bibitem[Navarro \em{et~al.}(1996)Navarro, Frenk, and White]{nfw1996}
Navarro, J.F.; Frenk, C.S.; White, S.D.M.
\newblock {The Structure of cold dark matter halos}.
\newblock {\em Astrophys. J.} {\bf 1996}, {\em 462},~563--575,
\newblock
  doi:{\changeurlcolor{black}\href{https://doi.org/10.1086/177173}{\detokenize{10.1086/177173}}}.

\bibitem[Burkert(1996)]{Burkert:1995}
Burkert, A.
\newblock {The Structure of dark matter halos in dwarf galaxies}.
\newblock {\em IAU Symp.} {\bf 1996}, {\em 171},~175,
\newblock 
  doi:{\changeurlcolor{black}\href{https://doi.org/10.1086/309560}{\detokenize{10.1086/309560}}}.

\bibitem[J.(1968)]{einasto1968}
Einasto, J.
\newblock On Galactic Descriptive Functions.
\newblock {\em Publ. Tartuskoj Astrofizica Obs.} {\bf
  1968}, {\em 36},~414.

\bibitem[Walker \em{et~al.}(2009)Walker, Mateo, Olszewski, Peñarrubia, Evans,
  and Gilmore]{walker2009}
Walker, M.G.; Mateo, M.; Olszewski, E.W.; Peñarrubia, J.; Evans, N.W.;
  Gilmore, G.
\newblock A Universal Mass Profile for Dwarf Spheroidal Galaxies?
\newblock {\em  Astrophys. J.} {\bf 2009}, {\em 704},~1274.

\bibitem[Adams \em{et~al.}(2014)Adams et~al.]{adams2014}
Adams, J.J.; others.
\newblock Dwarf Galaxy Dark Matter Density Profiles Inferred from Stellar and
  Gas Kinematics.
\newblock {\em  Astrophys. J.} {\bf 2014}, {\em 789},~63.

\bibitem[Colafrancesco \em{et~al.}(2006)Colafrancesco, Profumo, and
  Ullio]{Colafrancesco2006}
Colafrancesco, S.; Profumo, S.; Ullio, P.
\newblock Multi-frequency analysis of neutralino dark matter annihilations in
  the Coma cluster.
\newblock {\em Astron. Astrophys.} {\bf 2006}, {\em 455},~21--43.

\bibitem[Colafrancesco \em{et~al.}(2015)Colafrancesco, Marchegiani, and
  Beck]{gsp2015}
Colafrancesco, S.; Marchegiani, P.; Beck, G.
\newblock Evolution of Dark Matter Halos and their Radio Emissions.
\newblock {\em JCAP} {\bf 2015}, 
doi:10.1088/1475-7516/2015/02/032.

\bibitem[Evoli \em{et~al.}(2008)Evoli, Gaggero, Grasso, and Maccione]{dragon}
Evoli, C.; Gaggero, D.; Grasso, D.; Maccione, L.
\newblock Cosmic ray nuclei, antiprotons and gamma rays in the galaxy: A new
  diffusion model.
\newblock {\em J. Cosmol. Astropart.  Phys.} {\bf 2008}, {\em
  2008},~018.

\bibitem[Baltz and Edsj\"o(1998)]{baltz1999}
Baltz, E.A.; Edsj\"o, J.
\newblock Positron propagation and fluxes from neutralino annihilation in the
  halo.
\newblock {\em Phys. Rev. D} {\bf 1998}, {\em 59},~023511,
\newblock
  doi:{\changeurlcolor{black}\href{https://doi.org/10.1103/PhysRevD.59.023511}{\detokenize{10.1103/PhysRevD.59.023511}}}.

\bibitem[Baltz and Wai(2004)]{baltz2004}
Baltz, E.A.; Wai, L.
\newblock Diffuse inverse Compton and synchrotron emission from dark matter
  annihilations in galactic satellites.
\newblock {\em Phys. Rev. D} {\bf 2004}, {\em 70},~023512,
\newblock
  doi:{\changeurlcolor{black}\href{https://doi.org/10.1103/PhysRevD.70.023512}{\detokenize{10.1103/PhysRevD.70.023512}}}.

\bibitem[Colafrancesco and Blasi(1998)]{Colafrancesco1998}
Colafrancesco, S.; Blasi, S.
\newblock Clusters of Galaxies and the Diffuse Gamma Ray Background.
\newblock {\em Astropart. Phys.} {\bf 1998}, {\em 9},~227.

\bibitem[Longair(1994)]{longair1994}
Longair, M.S.
\newblock {\em High Energy Astrophysics}; Cambridge University Press: {Cambridge, UK}, 1994. 

\bibitem[Jeltema and Profumo(2008)]{jeltema2008}
Jeltema, .E.; Profumo, S.
\newblock Searching for Dark Matter with X-ray Observations of Local Dwarf
  Galaxies.
\newblock {\em Astrophys. J.} {\bf 2008}, {\em 686},~1045.

\bibitem[R.J. \em{et~al.}(1995)R.J., P.J., and M.C.H.]{miriad}
Sault, R.J.; Teuben, P.J.; Wright, M.C.
\newblock A retrospective view of Miriad. In {\em Astronomical Society of the
  Pacific Conference Series Vol. 77, Astronomical Data Analysis Software and
  Systems IV}; Shaw, R.A., Payne, H.E., Hayes,~J.J.E., Eds.;  Astronomical Society of the
  Pacific:{ San Francisco, CA, USA,} 1995; p. 433. 

\bibitem[Bertin and Arnouts(1996)]{sextractor}
Bertin, E.; Arnouts, S.
\newblock {SExtractor: Software for source extraction}.
\newblock {\em Astron. Astrophys. Suppl. Ser.} {\bf 1996}, {\em 117},~393--404,
\newblock
  doi:{\changeurlcolor{black}\href{https://doi.org/10.1051/aas:1996164}{\detokenize{10.1051/aas:1996164}}}.

\bibitem[Prandoni \em{et~al.}(2000)Prandoni et~al.]{prandoni2000a}
Prandoni, I.; Gregorini, L.; Parma, P.; De Ruiter, H.R.; Vettolani, G.; Wieringa, M.H.; Ekers, R.D.
\newblock The~ATESP radio survey I. Survey description, observations and data
  reduction.
\newblock {\em Astron. Astrophys. Suppl.} {\bf 2000}, {\em 146},~41--55.

\bibitem[Foley \em{et~al.}(2016)Foley et~al.]{kat7}
Foley, A.R.; Alberts, T.; Armstrong, R.P.; Barta, A.; Bauermeister, E.F.; Bester, H.; Blose, S.; Booth, R.S.; Botha,~D.H.; Buchner, S.J.; et al.
\newblock Engineering and science highlights of the KAT-7 radio telescope.
\newblock {\em Mon. Not. R. Astron. Soc.} {\bf 2016},
  {\em 460},~1664--1679,
\newblock
  doi:{\changeurlcolor{black}\href{https://doi.org/10.1093/mnras/stw1040}{\detokenize{10.1093/mnras/stw1040}}}.

\bibitem[Bonnivard \em{et~al.}(2015)Bonnivard, Combet, Maurin, Geringer-Sameth,
  Koushiappas, Walker, Mateo, Olszewski, and Bailey~III]{Bonnivard:2015}
Bonnivard, V.; Combet, C.; Maurin, D.; Geringer-Sameth, A.; Koushiappas, S.M.;
  Walker, M.G.; Mateo, M.; Olszewski, E.W.; Bailey, J.I., III.
\newblock {Dark matter annihilation and decay profiles for the Reticulum II
  dwarf spheroidal galaxy}.
\newblock {\em Astrophys. J.} {\bf 2015}, {\em 808},~L36,
\newblock
  doi:{\changeurlcolor{black}\href{https://doi.org/10.1088/2041-8205/808/2/L36}{\detokenize{10.1088/2041-8205/808/2/L36}}}.

\bibitem[Baring \em{et~al.}(2016)Baring, Ghosh, Queiroz, and
  Sinha]{fermi-decay}
Baring, M.G.; Ghosh, T.; Queiroz, F.S.; Sinha, K.
\newblock New limits on the dark matter lifetime from dwarf spheroidal galaxies
  using Fermi-LAT.
\newblock {\em Phys. Rev. D} {\bf 2016}, {\em 93},~103009,
\newblock
  doi:{\changeurlcolor{black}\href{https://doi.org/10.1103/PhysRevD.93.103009}{\detokenize{10.1103/PhysRevD.93.103009}}}.

\bibitem[Leite \em{et~al.}(2016)Leite, Reuben, Sigl, Tytgat, and
  Vollmann]{leite2016}
Leite, N.; Reuben, R.; Sigl, G.; Tytgat, M.; Vollmann, M.
\newblock Synchrotron emission from dark matter in galactic subhalos. A look
  into the Smith cloud.
\newblock {\em J. Cosmol. Astropart.  Phys.} {\bf 2016}, {\em
  2016},~021.

\bibitem[Storm \em{et~al.}(2017)Storm, Jeltema, Splettstoesser, and
  Profumo]{storm2017}
Storm, E.; Jeltema, T.E.; Splettstoesser, M.; Profumo, S.
\newblock Synchrotron Emission from Dark Matter Annihilation: Predictions for
  Constraints from Non-detections of Galaxy Clusters with New Radio Surveys.
\newblock {\em  Astrophys. J.} {\bf 2017}, {\em 839},~33.

\bibitem[Norris \em{et~al.}(2011)Norris, Hopkins, Afonso, Brown, Condon, Dunne,
  Feain, Hollow, Jarvis, Johnston-Hollitt, and et~al.]{emu}
Norris, R.P.; Hopkins, A.M.; Afonso, J.; Brown, S.; Condon, J.J.; Dunne, L.;
  Feain, I.; Hollow, R.; Jarvis, M.; Johnston-Hollitt, M.; et~al.
\newblock EMU: Evolutionary Map of the Universe.
\newblock {\em Publ. Astron. Soc. Aust.} {\bf
  2011}, {\em 28},~215–248,
\newblock
  doi:{\changeurlcolor{black}\href{https://doi.org/10.1071/AS11021}{\detokenize{10.1071/AS11021}}}.

\bibitem[Abbott \em{et~al.}(2016)Abbott et~al.]{des}
Abbott, T.; Abdalla, F.B.; Aleksić, J.; Allam, S.; Amara, A.; Bacon, D.; Balbinot, E.; Banerji, M.; Bechtol, K.; Benoit-Lévy, A.; et al.
\newblock The Dark Energy Survey: More than dark energy---An overview.
\newblock {\em Mon. Not. R. Astron.~Soc.} {\bf 2016},
  {\em 460},~1270--1299.
\newblock
  doi:{\changeurlcolor{black}\href{https://doi.org/10.1093/mnras/stw641}{\detokenize{10.1093/mnras/stw641}}}.

\bibitem[Keller \em{et~al.}(2007)Keller et~al.]{skymap}
Keller, S.C.; Schmidt, B.P.; Bessell, M.S.; Conroy, P.G.; Francis, P.; Granlund, A.; Kowald, E.; Oates, A.P.; Martin-Jones, T.; Preston, T.; et al.
\newblock {SkyMapper and the Southern Sky Survey}.
\newblock {\em Publ. Astron. Soc. Aust.} {\bf 2007}, {\em 24},~1--12,
\newblock
  doi:{\changeurlcolor{black}\href{https://doi.org/10.1071/AS07001}{\detokenize{10.1071/AS07001}}}.

\bibitem[Abdell \em{et~al.}(2009)Abdell et~al.]{lsst}
Abell, P.A.; Burke, D.L.; Hamuy, M.; Nordby, M.; Axelrod, T.S.; Monet, D.; Vrsnak, B.; Thorman, P.; Ballantyne,~D.R.; Simon, J.D.; et al.
\newblock LSST Science Book, Version 2.0.
\newblock {\em arXiv} {\bf 2009}, { arXiv:0912.0201}.

\end{thebibliography}



\end{document}